# Electric Field Induced Agglomeration of Nearly Ferroelectric Superconducting YBa$_2$Cu$_3$O$_{7-\delta}$ Powders

B. Andrzejewski[a*], W. Bednarski[a], K. Chybczyńska[a], A. Hilczer[a], M. Matczak[b], F. Matelski[c]

[a] Institute of Molecular Physics, Polish Academy of Sciences
Smoluchowskiego 17, PL-60179 Poznań, Poland
* corresponding author: andrzejewski@ifmpan.poznan.pl

[b] NanoBioMedical Centre, Adam Mickiewicz University
Umultowska 85, PL-61614 Poznań, Poland

[c] Poznań University of Technology, Nieszawska 13A, PL-60965 Poznań, Poland

*Abstract* — **We report on the effect of agglomeration forced by strong electric field in fine particles of nearly ferroelectric YBa$_2$Cu$_3$O$_{7-\delta}$ superconductor. It turns out that the particles from agglomerates exhibit different morphology than the rest of powder that attaches to high-voltage electrodes. Study by means of electron paramagnetic resonance revealed in the powder attached to electrodes a narrow spectrum superimposed on Cu$^{2+}$ anisotropic spectrum common for YBa$_2$Cu$_3$O$_{7-\delta}$ superconductors. We assume that this narrow spectrum originates from nanopolar regions generated by strong electric discharges taking place during the experiment. Consequently, the effect of agglomeration can be explained in terms of electrostatic interactions between the particles containing nanopolar regions with strong electric dipolar moments.**

## I. INTRODUCTION

More than ten years ago, Tao et al. [1] reported that a fraction of fine superconducting particles agglomerate into small balls in enough strong electric field. The rest of the superconducting powder attaches to the electrodes. These agglomerates are usually composed of numerous particles compacted so strong that they can survive rapid collisions with electrodes during fast oscillations. The bouncing of superconducting agglomerates is forced by electrostatic interactions because agglomerates become charged after contacting with electrodes and during oscillations they transport electric charge between electrodes. The size of superconducting balls depends on the electric field i.e. big agglomerates appear above some threshold field, next their size decreases as the strength of the electric field increases and the agglomerates disappear above the second threshold field. The agglomerates arise in the superconducting state only and they decay exactly above the transition to normal state. In normal state, superconducting powders behave like metallic micro particles *i.e.* they form conducting paths that short-circuit high-voltage electrodes.

The "Tao effect" seems to be very common both for high-temperature and conventional superconductors. Actually, it was observed for Bi$_2$Sr$_2$CaCu$_2$O$_{8+x}$, NdBa$_2$Cu$_3$O$_x$, YbBa$_2$Cu$_3$O$_x$ and YBa$_2$Cu$_3$O$_{7-\delta}$ (YBCO) high-temperature superconductors (HTSC) at liquid nitrogen temperature [1], for MgB$_2$ superconductor at moderate temperatures [2] and also for conventional, metallic superconductors like Pb, V, V$_3$Ga, NbN and Nb$_3$Sn at temperature of liquid helium [3]. The agglomeration of the superconducting particles take place regardless the superconductor is type I (V, Pb) or type II superconductor (HTSC, MgB$_2$, V$_3$Ga, NbN, Nb$_3$Sn). The Tao effect depends also on the frequency of electric field and it was observed that the superconducting balls formed columns or chains connecting electrodes [4]. This behavior is similar to the electrorheological fluids and demonstrates that the superconducting particles are also polarized in electric field.

The explanations of „Tao effect" assume Josephson coupling energy between superconducting Cu-O planes in high-temperature superconductors [1], positive surface energy [3] or charge expulsion from superconductors based on unconventional theory of „hole superconductivity" [5-7]. However, none of these theories is able to explain all experimental results in satisfactory way.

Recently Gosh et al. [8] reported on important doubts concerning the origin of the "Tao effect" because the effect of agglomeration appeared also for insulating fine particles immersed in liquid nitrogen. He concluded that this effect is trivial and not related to superconductivity. According to Hirsch, fine particles have tendency to agglomerate because of humidity which serve as "glue" for micro particles. The water can be absorbed by superconducting particles but it can also condense in liquid nitrogen.

To shed some light on the origin of the "Tao effect" we decided to study morphology and also the composition of the fraction of superconducting powder which forms the agglomerates and the fraction which attaches to the electrodes. We found that these two powders exhibit different shapes and properties.

## II. EXPERIMENTAL

FEI NovaNanoSEM 230 scanning electron microscope (SEM) was applied to study the details of morphology of YBCO superconducting grains. Electron paramagnetic resonance studies (EPR) were performed with a Bruker ElexSys E500 spectrometer operating in the X-band (~9.5 GHz). EPR spectra were recorded as the first derivative of the microwave power absorption vs. magnetic field. The magnetic field was modulated with a frequency of 100 kHz.

The YBCO powder used in this experiment was synthesized by Superconductive Components Inc. The purity of $YBa_2Cu_3O_{7-\delta}$ compound was 99.99% with the oxygen index $\delta$ close to 0.1 and the mean diameter of the particles about 2 μm. Before the experiment, the YBCO powder was dried in VD23 Binder GmbH dryer during about 2 h at rough vacuum to remove any humidity which could facilitate formation of the aggregates. Next the powder was immediately dropped to the container filled with liquid nitrogen. Only fresh liquid nitrogen *i.e.* without any ice and oxygen pollutions was used in the experiment. To estimate the rate of water absorption, the dry YBCO powder was exposed to the air with humidity about 90% and an increase in powder mass was measured using Radwag Inc. balance with the accuracy 1μg.

The "Tao effect" was studied by means of a horizontal capacitor setup. A pair of parallel aluminum electrodes used for this capacitor had relatively large size; 60 mm in length and 30 mm in height, which provided uniform electric field. The electrodes were separated by 8 mm thick Teflon spacer and fitted to polystyrene container filled with liquid nitrogen.

## III. RESULTS AND DISCUSSION

The relative increase in mass of dry YBCO powder caused by water absorption is presented in Fig. 1. Even after 1 h of the exposition to air the change of mass didn't exceeded 0.1%. Due to the short time of transferring the powder from drier to the experimental set-up (about 3-4 min) the amount of absorbed water was negligible and it couldn't be responsible for the agglomeration of YBCO powder.

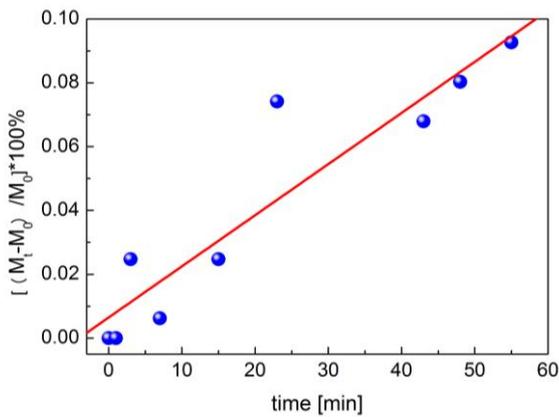

Figure 1. The mass increase of YBCO powder due to water absorption.

Formation of the agglomerates of superconducting YBCO particles immersed in liquid nitrogen started at the field about 7.5 kV/cm and was preceded by multiple electric discharges between the high-voltage electrodes. Usually, only 2-3 agglomerates were formed during one experiment and bounced in the capacitor whereas the rest of the powder was attached to the electrodes. This effect was thus similar to the behavior reported earlier by Tao et al. [1]. Also, it turned out that some agglomerates were compacted so hard that they survived above the critical temperature and could be taken out of the container with liquid nitrogen. The agglomerates were stable and they didn't decay even after one month. The assembly of the agglomerates obtained in this way are shown in Fig. 1.

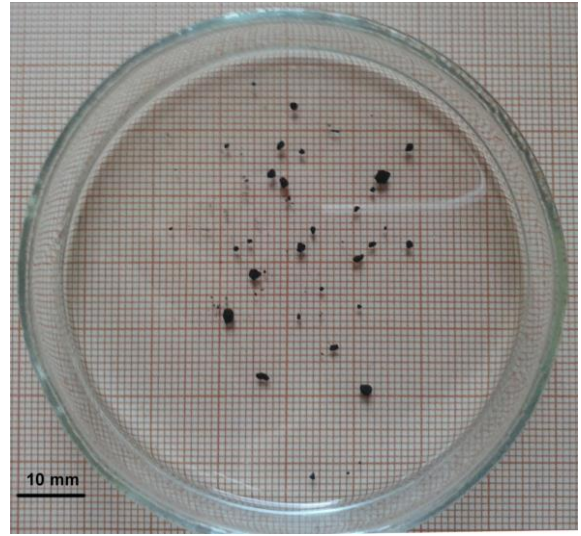

Figure 2. The assembly of YBCO agglomerates obtained after several experiments.

The largest agglomerates exhibited diameter nearly 2 mm whereas their mean size was about 0.7 mm. The histogram of sizes of YBCO agglomerates is shown in Fig. 3.

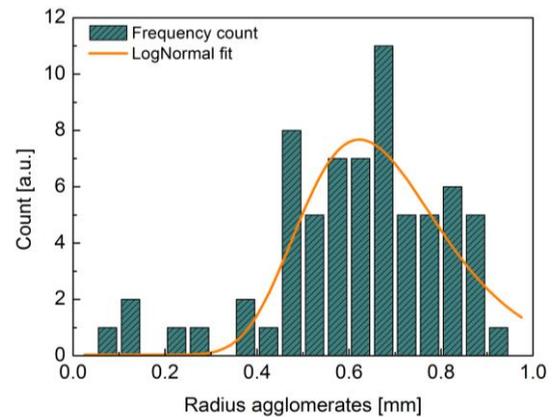

Figure 3. The histogram of sizes of YBCO agglomerates.

The size distribution of the agglomerates was fitted with log-normal function commonly used to describe granular random systems:

$$f(D) = \frac{1}{\sqrt{2\pi\sigma^2}} \frac{1}{D} \exp\left(-\frac{1}{2\sigma^2} \ln^2\left(\frac{D}{\langle D \rangle}\right)\right) \quad (1)$$

where $\langle D \rangle$ denotes the median radius of the nanoparticles and $\sigma$ is the distribution width. The best fit of eq. (1) to the data presented in the histogram Fig. 3 was obtained for the median radius $\langle D \rangle$=0.66 mm and standard deviation $\sigma$=0.24.

An example of the big ball-like agglomerate, composed of numerous YBCO micro particles (~$10^6$) is presented in SEM micrograph in Fig. 4.

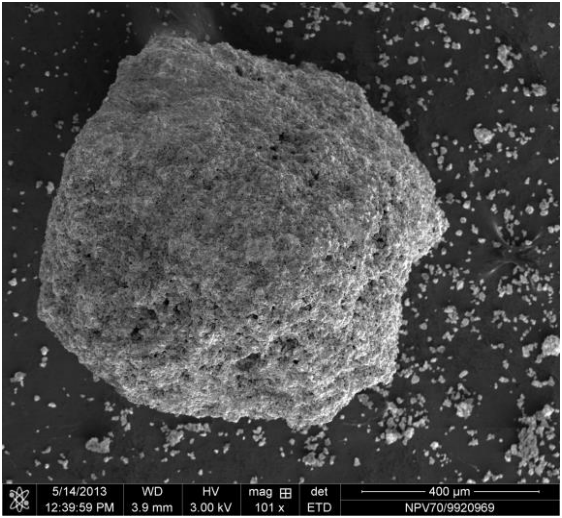

Figure 4. The SEM micrograph of YBCO agglomerate.

The details of this agglomerate surface are shown in Fig. 5a and can be compared with SEM micrograph of the powder that attached to high-voltage electrodes presented in Fig. 5b. The powder at the surface is highly compacted whereas the loosely grains are located mainly inside the agglomerates.

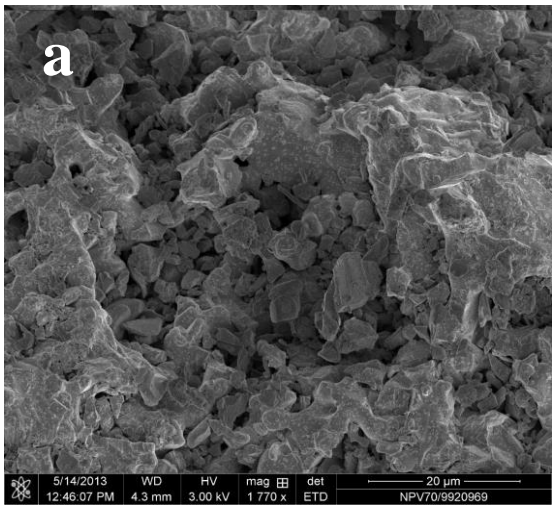

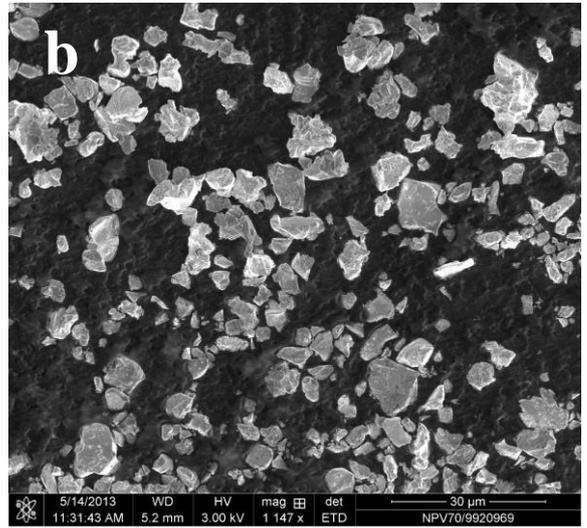

Figure 5. The SEM micrograph presenting the details of surface of YBCO agglomerate a) and the powder collected at high-voltage electrodes b).

The SEM study of these two fractions of superconducting powders revealed also different sizes of the particles that compose the agglomerates and the particles that attach to the electrodes. Namely, the median diameter of the particles in agglomerates determined from the log-normal fit eq. (1) of the histogram shown in Fig. 6 is about $\langle D \rangle \approx 1.6$ μm whereas the size of the particles from electrodes is larger and equal $\langle D \rangle \approx 2.4$ μm.

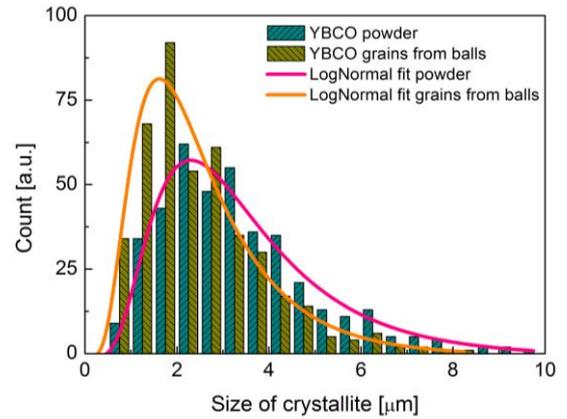

Figure 6. The histogram of diameters of nanoparticles from the agglomerates and high-voltage electrodes. The lines are the best fits of the log-normal distribution eq. (1) to the data.

The particles are different not only with respect to the size but also they exhibit different shape represented by eccentricity $e=(1-a^2/b^2)^{1/2}$ where $a$ and $b$ are the short and long axis of the grain (if the grains are approximated by the ellipsoids of revolutions). The results of measurements are shown in histograms in Fig. 7. The distribution of particle eccentricity can be fitted by means of the modified Rayleigh probability function [9]:

$$f(e) = \frac{1-e}{\sigma^2} \exp\left(-\frac{(1-e)^2}{2\sigma^2}\right) \qquad (2)$$

where $\sigma$ is related to the mean eccentricity $\langle e \rangle$ of the particles trough $1-\langle e \rangle = (\pi/2)^{1/2}\sigma$. The best fit of eq. (2) to the data in the histogram Fig. 7 was obtained for the mean eccentricity $\langle e \rangle = 0.67(1)$ in the case of the particles from agglomerates and $\langle e \rangle = 0.75(1)$ for the powder collected from the electrodes. Therefore the particles that compose the agglomerates are more spherical than the particles that have tendency to attach to the electrodes. Larger size and more elongate shape of these particles should increase their electric dipolar moment, especially if they are charged or if they include some dielectric or ferroelectric insulating regions. The presence of polar regions should facilitate attraction of these superconducting particles to the high-voltage electrodes.

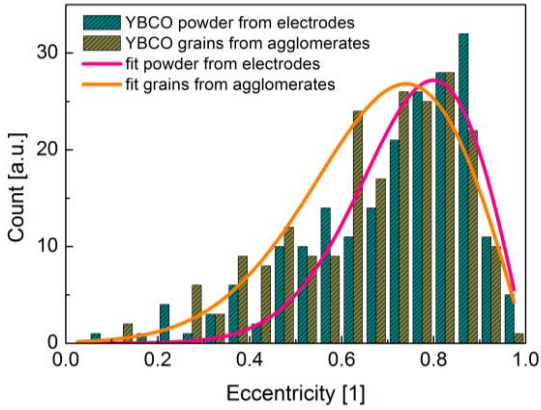

Figure 7. The histograms of eccentricity of nanoparticles collected from the agglomerates and from the electrodes. The lines are the best fits of the function eq. (2) to the data.

The results obtained from the measurements of EPR spectra are another strong argument that speak in favor of the differences between these two fractions of superconducting powders. The EPR spectrum recorded for the particles from agglomerates presented in Fig. 8a is common for YBCO superconductors [10]. It results from $Cu^{2+}$ paramagnetic centers located in the lattice sites with axial symmetry. The parallel and perpendicular g-factors of this wide line are $g_\parallel = 2.223$ and $g_\perp = 2.046$, respectively. The YBCO powder from the electrodes exhibits similar $Cu^{2+}$ EPR spectrum with g-factors $g_\parallel = 2.226$ and $g_\perp = 2.047$ but also superimposed, additional very narrow line with $g = 2.051$ (see Fig. 8b). Moreover, this narrow line disappears almost completely during *c.a.* two weeks after the experiment was performed and only wide $Cu^{2+}$ line remains. It is therefore reasonable to assume that the narrow line originates from the paramagnetic centers created during the experiment, most probable during multiple electric discharges observed during initial stage of the experiment.

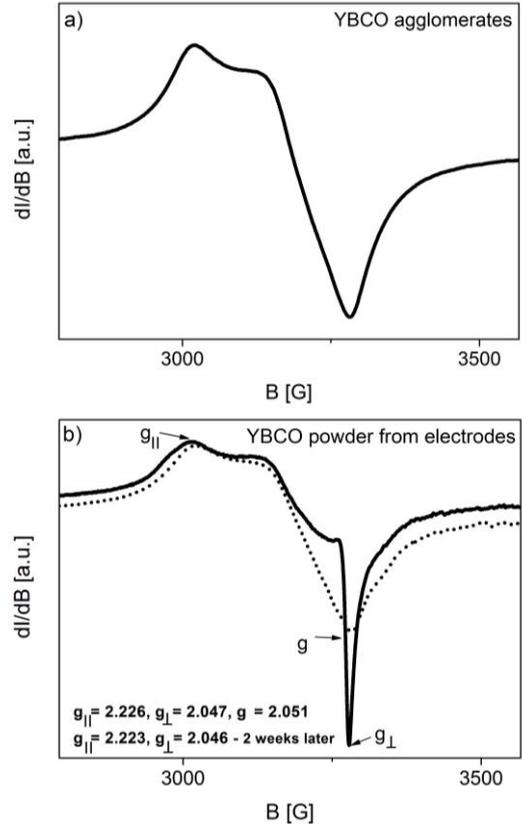

Figure 8. The EPR spectra for the YBCO powder from agglomerates a) and from the electrodes recorded immediately after the experiment (solid line) and two weeks later (dot line) b).

The strong electric discharges mediated by the powder between the electrodes generate local heat and high temperature, mainly at the intergranular contacts. This in turn can lead to loss of oxygen in some regions of YBCO particles. If the oxygen deficiency in these regions increases above $\delta = 0.65$ they are no longer superconducting and convert to insulating ones [11]. High-temperature YBCO superconductor belongs to the family of nearly ferroelectric superconductors [12-14] and therefore these insulating regions can exhibit some spontaneous or induced electric dipolar moment. If the electrostatic force due to the electric field from electrodes dominates over interparticle interactions then rich in polar regions particles easily attach to the electrodes. Also, larger size and eccentricity of these particles facilitate this effect. On the other hand, smaller and more spherical grains, which are close one to the other and do not contain many polar regions are not attracted enough strong to the electrodes and they can aggregate due to interparticle electrostatic forces. Obviously, the amount of powder that satisfies above requirement is rather small and this explains why only a few agglomerates are usually created during single experiment.

## IV. CONCLUSIONS

We have observed the Tao effect *i.e.* agglomeration of superconducting fine particles induced by a strong electric field in dry YBCO powder. This excludes origin of the Tao effect due to the absorbed water which can serve as glue for fine particles. The study of morphology of superconducting particles reveal substantial differences in the size and shape of the YBCO powder that form the agglomerates and the powder that attach to high-voltage electrodes. Namely, the YBCO grains from the agglomerates are smaller and more spherical than the elongated and large grains collected on the electrodes. These two fractions of powders exhibit also some differences in composition investigated by EPR method. The EPR spectrum for the grains from agglomerates results from $Cu^{2+}$ paramagnetic centers located in the lattice sites with axial symmetry and is common for YBCO superconductor. For the YBCO powder from the electrodes there is observed additional very narrow line. We assume that this line originates from paramagnetic centers in the oxygen deficient regions created during strong electric discharges in the particles between the electrodes. These regions can be insulting and thus can exhibit some spontaneous or induced electric dipolar moment. Therefore, the agglomeration of superconducting powders can be explained in terms electrostatic interactions between the grains with dipolar moments. These particles which are more elongate and rich in polar regions do not agglomerate but attach to the high-voltage electrodes.

## V. ACKNOWLEDGEMENTS

This project has been supported by National Science Centre under the project No. N N507 229040. M.M. was supported through the European Union - European Social Fund and Human Capital - National Cohesion Strategy.